\documentclass[lettersize,journal]{IEEEtran}
\usepackage{amsmath,amsfonts}
\usepackage{algorithmic}
\usepackage{algorithm}
\usepackage{array}
\usepackage[caption=false,font=normalsize,labelfont=sf,textfont=sf]{subfig}
\usepackage{textcomp}
\usepackage{stfloats}
\usepackage{url}
\usepackage{verbatim}
\usepackage{graphicx}
\usepackage{cite}
\usepackage[nolist,nohyperlinks]{acronym}
\usepackage{comment}
\usepackage{multirow}
\usepackage{subfig}
\usepackage{xcolor,soul}
\begin{document}




\title{Distributing Intelligence in 6G Programmable Data Planes for Effective In-Network Intrusion Prevention 
}


\author{Mattia G. Spina, Floriano De Rango, Edoardo Scalzo, Francesca Guerriero, Antonio Iera 

\thanks{M. Spina, F. De Rango, and A.Iera are with Dept. DIMES, University of Calabria, Italy and CNIT, Italy.}\thanks{F. Guerriero and E. Scalzo are with Dept. DIMEG, University of Calabria.}\thanks{This work was partially supported by the European Union under the Italian National Recovery and Resilience Plan (NRRP) of NextGenerationEU, partnership on ``Telecommunications of the Future” (PE00000001 - program ``RESTART”). This work has been submitted and accepted to the IEEE for publication. Copyright may be transferred without notice, after which this version may no longer be accessible.}
}

\markboth{}%
{Shell \MakeLowercase{\textit{et al.}}: A Sample Article Using IEEEtran.cls for IEEE Journals}


\maketitle

\begin{abstract}
The problem of attacks on new generation network infrastructures is becoming increasingly relevant, given the widening of the attack surface of these networks resulting from the greater number of devices that will access them in the future (sensors, actuators, vehicles, household appliances, etc.). Approaches to the design of intrusion detection systems must evolve and go beyond the traditional concept of perimeter control to build on new paradigms that exploit the typical characteristics of future 5G and 6G networks, such as in-network computing and intelligent programmable data planes. The aim of this research is to propose a disruptive paradigm in which devices in a typical data plane of a future programmable network have 
anomaly detection capabilities and cooperate in a fully distributed fashion to act as an ML-enabled Intrusion Prevention System ``embedded" into the network. The reported proof-of-concept experiments demonstrate that the proposed paradigm allows working effectively and with a good level of precision while occupying overall less CPU and RAM resources of the devices involved. 
\end{abstract}


\section{Introduction}
\IEEEPARstart{O}{ne} of the key aspects characterizing the transition from previous generation systems to 5G and to the imminent 6G is the ability to manage complex data flows of various kinds, originating from a multitude of devices. 
This brings with it a growing threat represented by the exponential increase and diversification of network access points, which implies a more extensive attack surface. Future networks will therefore be increasingly exposed to external attacks, which in turn are becoming more and more sophisticated \cite{Park23}, making the need to design equally sophisticated network protection mechanisms extremely pressing.
Among these are Network Intrusion Detection Systems (IDS), which can be classified as passive, when they detect and record malicious behavior, or active, when they also take actions such as dropping packets or blocking Source IP when an attack is detected. The latter is also known as Intrusion Prevention System (IPS).

P4 \cite{bosshart2014p4} is one of the high-level languages available for programming network data plane devices, which is gaining momentum in research and industry due to its feature of being "OpenFlow independent" and the possibility it offers to program different types of network devices, switches and NICs (e.g., fixed-function ASICs, NPUs, and FPGAs) \cite{Cordero}.

In-Network Computing (INC) is another key enabler that has gained momentum recently, also due to the availability of programmable data plane (PDP) devices. In fact, it allows not only routers but also switches and Smart Network Interface Cards (smartNICs) to add to their typical packet forwarding capabilities also computational functionalities previously reserved only for Edge or Cloud servers \cite{Tarik23},\cite{Capone}.

Another feature to be exploited is the one that will perhaps differentiate 6G networks the most from their predecessors, namely the massive distribution of intelligence across their data plane components for effective network management.
The presence of Machine Learning (ML)-enabled Programmable Data Planes (PDPs) allows to embed the INC concept in 6G and offload application-specific tasks to data plane entities \cite{MAHM23}.

In this paper, in the wake of the mentioned innovations, the goal is to significantly contribute to the advancement of the state of the art of next-generation network intrusion detection techniques by proposing a disruptive solution that jointly exploits the concepts of \textit{Distributed AI} and \textit{Programmable Data Plane}.

Specifically, the main contribution does not simply consist in equipping elements of PDPs of future networks with pre-trained ML models (Decision Tree (DT), Random Forest (RF), etc.), to enable them to classify flows and packets \textit{while crossing the network}; this is a topic widely covered by valid studies already available.

Rather, we want to evaluate the potential of splitting a stronger ML model into weaker models distributed across groups of multiple PDP devices and exploiting their coordinated action to implement a fully distributed and cooperative mechanism to identify potential attacks and block them directly in-network from the first signs of their presence.

The key paradigm shift that this proposal aims to provide to overcome the current state of the art, consists in shifting the research focus on AI-based in-network IPS systems from the model encoding problem - sometimes strictly dependent on the limitations imposed by existing PDP programming languages and related hardware technologies - to a network management problem leveraging the joint combination of programmable networks and an AI model decomposition strategy.

Compared to what is available from the state of the art on the subject, the disruptive approach we propose consists in \textit{making the network data plane itself capable of countering network intrusion attempts} not only by classifying the traffic but also by applying countermeasures directly in-network. 

Even the alternative state-of-the-art in-network solution that concentrates SLs in single nodes shows the main weakness of losing effectiveness due to the excessively heavy load that it would add to the data plane devices in addition to that of the classic functions performed (routing \& forwarding).

Instead, our solution based on in-network chains of highly distributed weak (and lightweight) learners can be very accurate and timely in countering attacks and can reduce the additional load on network traffic, thus offering a valid solution to preserve network performance and ensure scalable detection times in case of increasing attack rates and volumes. 

Thanks to the improved scalability due to the reduced complexity of the AI model deployed on every single PDP and therefore the limited imposed overhead, the new distributed in-network IPS paradigm aims to block attacks by minimizing attack detection and response times, while maximizing security coverage and reducing the impact of IPS features on host device performance and traffic QoS, overcoming the limitations of the existing \textit{not scalable} solutions that strive to deploy models entirely on networking devices.

The remainder of the document is organized as follows. Section 2 describes some relevant works from the literature, while Section 3 introduces the reference functional architecture supporting the proposed paradigm. Section 4 reports a proof-of-concept testing campaign, while the main open research challenges to be addressed to fully develop the proposed solution and the conclusions are reported in Sections 5 and 6, respectively.

\section{Related Works}

A large body of literature is available on IDS/IPS solutions in software-defined networks that leverage ML/DL techniques.
In contrast, the development of IDS/IPS solutions inside programmable Data Planes is a topic that has only recently received due attention, also thanks to the availability of flexible languages to program PDP devices, such as P4, as summarized in the review paper \cite{Gao}.

Most solutions involve the implementation of ML models running entirely on individual network devices, primarily intended for network traffic classification purposes. Others avoid overloading the packet processing and forwarding capabilities by distributing in-network intelligence in programmable switches for both network telemetry \cite{saquetti2021toward} and network traffic classification \cite{xavier2021programmable} purposes.

An interesting work in the field of in-network inference, using DT and RF, is \cite{Busse}, where a mechanism is proposed to translate DT and their RF ensemble inside programmable switches. In \cite{Zheng}, a framework for encoding different ML models on PDPs and a strategy for mapping the model into rules to be installed on switches for easier update are proposed.

The authors of \cite{Lee2020Nov} and \cite{NetBeacon} also proposed a framework that encodes a specific RF model on a PDP, without considering any dependencies or cooperation between switches. The model (RF) is entirely encoded inside a single switch that runs its own inference.

The ones reported are just a few examples among many available that, due to space limitations, cannot be listed here. However, none of those available in the literature propose in-network IDS/IPS solutions that are \textit{fully distributed and orchestrated inside a PDP}, like the one we propose.

By this, we mean that the detection tasks are performed inside the network through the cooperation of PDP devices enhanced with weak learning modules (DTs) without any additional involvement of the Control Plane. Instead, the cooperation is exclusively enabled through a custom packet header that carries within it the intermediate inferences and the information needed to perform the final detection through a Majority Voting that reconstructs the behavior of a complex SL. 

\section{Proposed Framework}
This section provides details on the key components and reference functional architecture of a Ubiquitous In-Network IPS that leverages a PDP enabled to support in-network distributed intelligence. 

\begin{figure*}[!h]
\centering
\includegraphics[width=17cm]{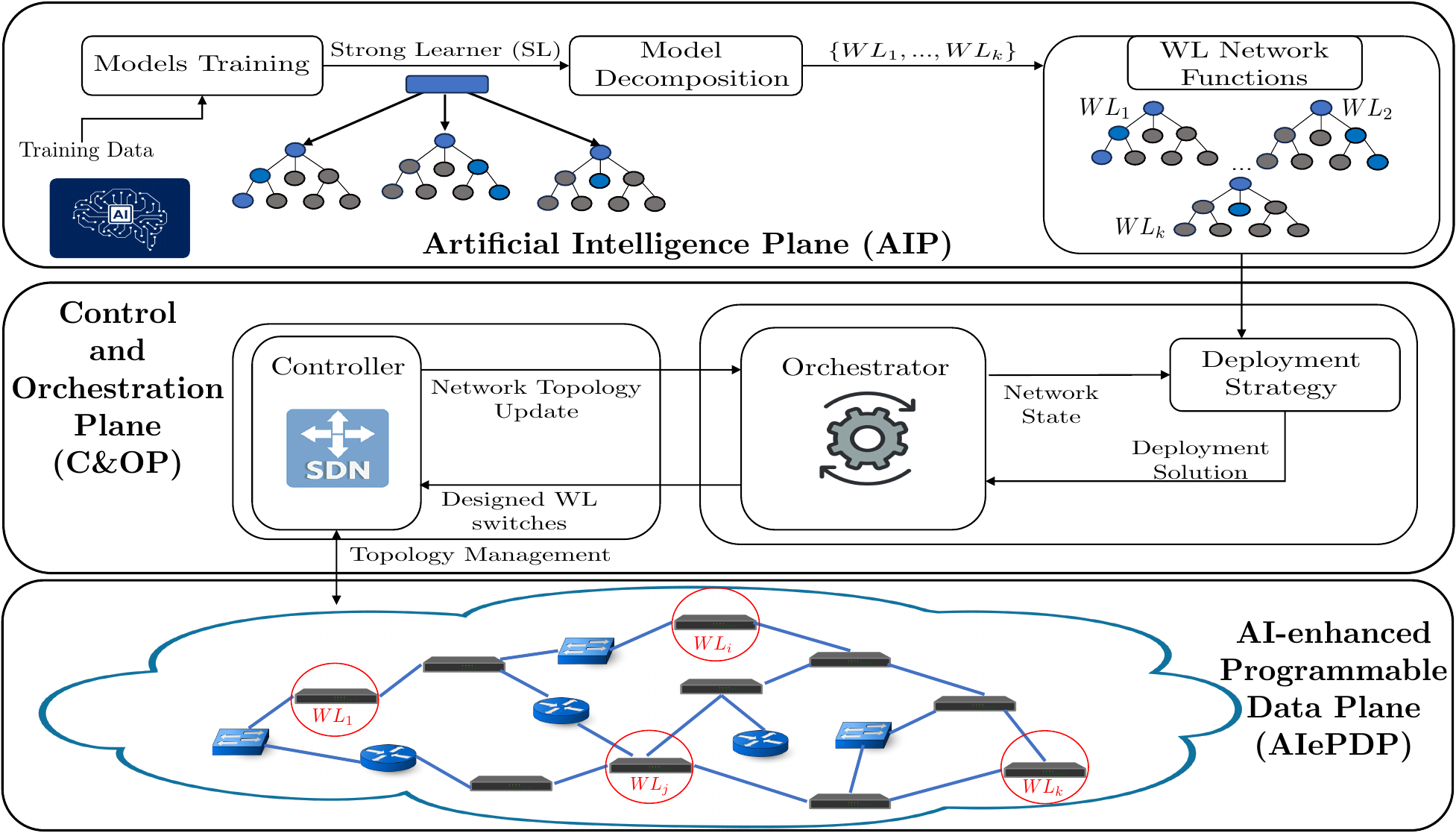}
\caption{Logical Architecture of the Proposed In-Network Computing based Active Intrusion Prevention System}
\label{fig:system_model}
\end{figure*}

Fig. \ref{fig:system_model} depicts the main framework components with relevant interactions.
More specifically, the framework is logically organized into three planes, \textbf{A}rtificial \textbf{I}ntelligence \textbf{P}lane (\textit{AIP}), \textbf{C}ontrol  \&  
 \textbf{O}rchestration \textbf{P}lane (\textit{C\&OP}), \textbf{AI}-\textbf{e}nhanced \textbf{P}rogrammable \textbf{D}ata \textbf{P}lane (\textit{AIePDP}). 
The main tasks associated with each Plane are detailed below.

\subsection*{Artificial Intelligence Plane (AIP)}
The AIP is responsible for ML/DL management-related operations. Specifically, it encompasses the environment devoted to model training, decomposition, and mapping onto Virtual Network Functions (VNFs). The first component, \textit{Model Training}, handles the training of a model and performs all the tasks necessary to achieve this goal.
After a good-quality set of data is obtained from the previous task, the next task is \textit{model-building}, which consists in designing, testing, and evaluating different DL/ML ensemble models by feeding them with the previously generated dataset. Considering different types of models (e.g., RFs, DTs etc.), a \textit{Model Fine-Tuning} step is performed by extensively searching within the hyperparameters space to find the parameter values of a model best suiting the context domain coded within the dataset. The output of this task is an SL or a set of SLs that are then processed by the next component, namely \textit{Model Decomposition}, which decomposes them into individual Weak Learners (WLs). Once WLs have been generated, they are encoded as WL-VNFs and made available to the underlying plane through the \textit{WL Network Functions} catalog.

\subsection*{Control \&  Orchestration Plane (C\&OP)}

The functions implemented in this Plane have the purpose of (i) developing the \textit{Deployment Strategy} of the WL-VNFs in the catalog, determining their optimal displacement across the PDP devices, and (ii) enabling the selected switches (by deploying on them the WL-VNFs) to operate together as an in-network IPS.

As for the first function, given the WL-VNFs catalog and the current snapshot of the network, the objective is to find the strategy that maximizes the \textit{security coverage} of the network. This means to find a set of WL-VNFs and relevant hosting switches that can effectively cover the considered network, intercepting the greatest possible number of flows to analyze (and on which to possibly intervene). Subsequently, the SDN controller takes the output from the orchestrator and implements the chosen WL-VNF deployment plan in the network.

The functions of this plane are crucial because the optimal deployment of such WL-VNFs allows to reduce the probability that an attack may not be detected and can reach its target.
Besides, network topology and traffic distribution may change over time, hosts could be shut down, links may fail, and data plane devices may go down.
Therefore, it must be possible to dynamically change the implemented deployment strategy to ensure that the \textit{security coverage} is preserved. For this reason, the SDN controller sends network snapshots to the WL-NFVs orchestrator to notify it of any network topology changes that require the computation of a new WL-VNF deployment plan.

\subsection*{AI-enhanced Programmable Data Plane (AIePDP)} 
This plane hosts both programmable and classic network forwarding elements (e.g. smart NICs, Programmable Switches, off-the-shelf hardware that hosts general-purpose VNFs, etc.). According to the WL-VNF \textit{deployment plan} some of them can be elected as in-network IPS elements that execute a WL-VNF function. Once a WL-VNF enabled PDP device detects a malicious flow, a cooperative and distributed analysis is started. The suspicious flow is marked as malicious and will be analyzed by the other on-the-path network elements equipped with the remaining WL-VNFs that need to be executed on the flow to re-construct the original SL. The network flow will cross the whole chain of WL-VNFs, by carrying along with it the produced intermediate inferences results encoded in the mentioned custom header. 
Specifically, for an SL consisting of $N$ WLs, the custom header is structured as

\begin{center}
 [$WL^{ID}_{1}|WL^{ID}_{2}|...|WL^{ID}_{N}|O_1|O_2|...|O_N$] 
\end{center}

where each $WL^{ID}_{i}$, with $i \in \{1, 2, ..., N\}$, denotes a $\lceil log_2(N) \rceil$-bit field encoding the unique identifier of the $i$-th WL. The fields $O_i$ (1-bit each) represent the intermediate results produced by the corresponding $WL^{ID}_{i}$ during the inference process.
The last WL-VNF receiving the flow produces its inference, verifies the existence of all the information needed to make a decision, and determines the final outcome about the flow by performing a Majority Voting. This is what we call ``\textit{Projecting the Ensemble Learning over the network}". Flows detected as malicious, are thus blocked by the cooperating in-network IPS elements \textit{without the need for human or controller intervention}. In such a way, the network is able to self-adapt to malicious events, counteracting them in an autonomous and timeliness way Obviously, enabling a limited-resource element, like a switch, to execute an ML/DL model without affecting its primary task, i.e. forwarding network flows, is a key issue to address.

\begin{figure}[!h]
    \centering
    \includegraphics[width=3in]{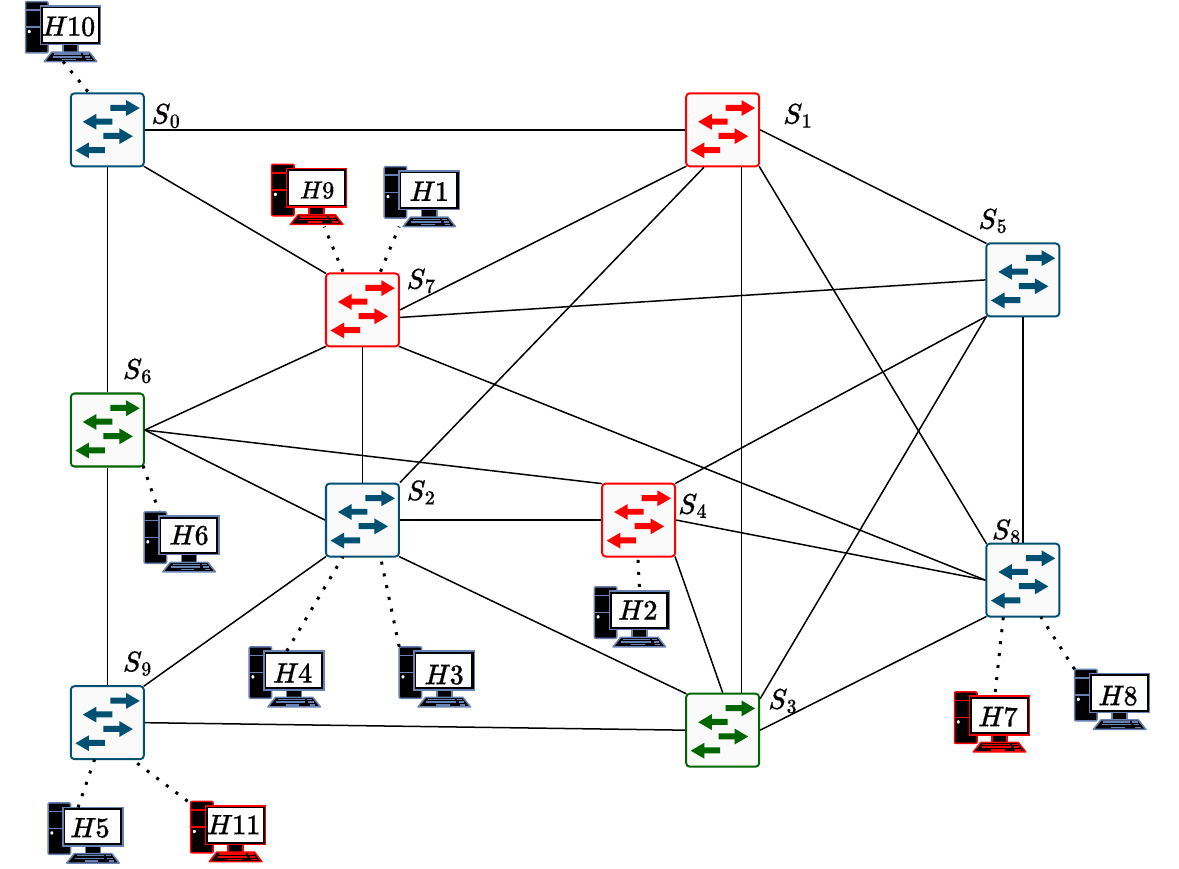}
    \caption{Considered topology and WL-VNF deployment strategy.}
    \label{fig:fig1}
\end{figure}

\begin{figure}[!h]
    \centering
    \includegraphics[width=3in]{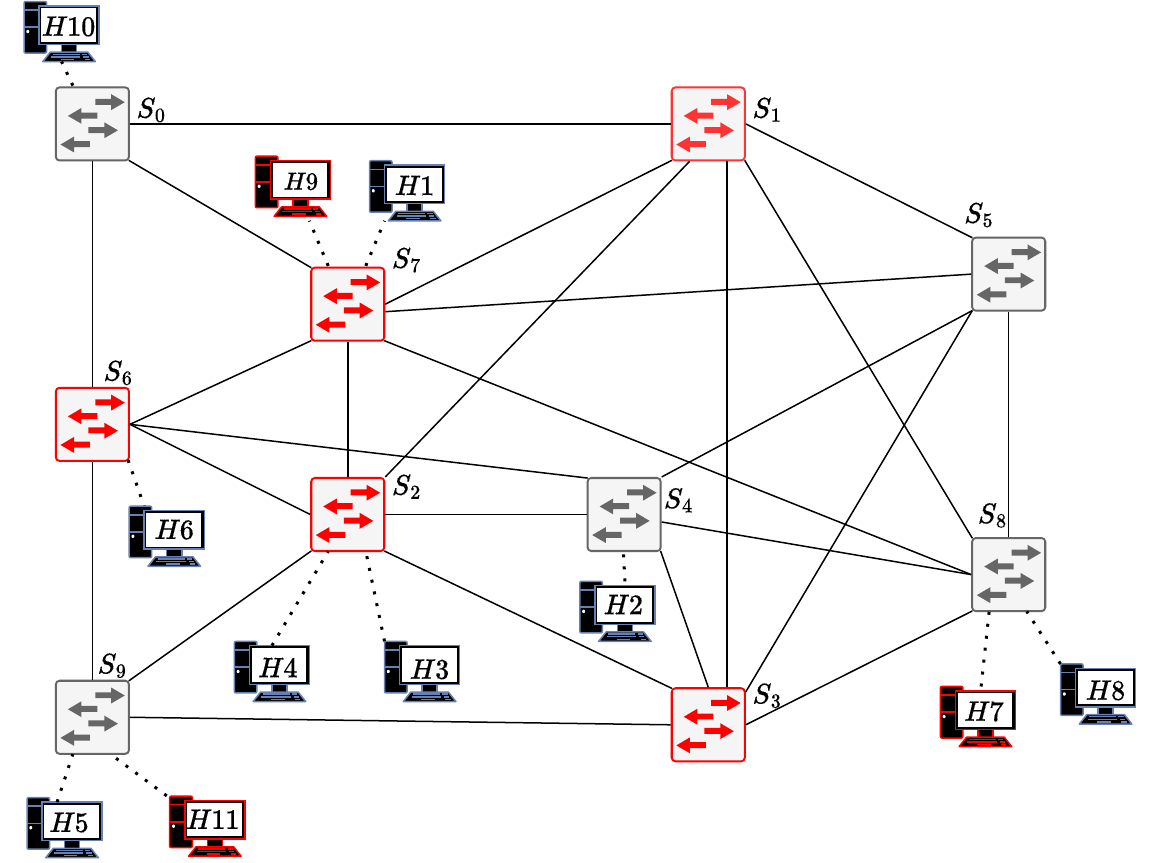}
    \caption{Considered topology and SL-VNF deployment strategy.}
    \label{fig:sl-top}
\end{figure}

\section{Performance Evaluation}

Below, we present the results of a test campaign providing an initial proof-of-concept for the proposed paradigm. The analysis focuses on its ability to operate with quality and effectiveness comparable to a benchmark solution (SL) while reducing the overhead on data plane devices tasked with detecting malicious activity.

\subsection{Experimental Setup}

To test the activities carried out at the \underline{AIP plane}, an RF composed of three DTs, each with a maximum depth of seven, is considered as a reference SL. The model has been trained and tested on the CSE-CIC-IDS2018 (available at https://registry.opendata.aws/cse-cic-ids2018) 
dataset with a train-test split ratio of 70\% and 30\% and a total number of features equal to 72. 
Traffic classification accuracy is assessed through a dedicated test. Benign and malicious flows from CIC-IDS-2018 are injected into the PDP device, ensuring test data is unseen during training. The PDP's results are then compared with the dataset's ground truth.

The obtained results achieve an accuracy of 94.54\%, precision of 0.8249, recall of 0.8249, and F1-score of 0.8985. False Positive Rate and False Negative Rate also show negligible avg. values of 0.056 and 0.013, respectively. These results are perfectly in line with the offline accuracy test of the model performed using the Python environment. 

The design of SL is carefully guided to balance accuracy and complexity through feature sampling. SL is trained on 72 features and then mapped to an SL-VNF.
It is then decomposed into its WL components, each of which is trained on 33\% of the total features, thus introducing diversity, reducing overfitting, and minimizing the correlation between WLs. After training, each WL is also mapped to a WL-VNF.
To run SL-VNF, all 72 features of a given flow need to be computed, while each WL-VNF only needs the 24 features it was trained on; this ensures faster training and model efficiency. 

The detection is triggered on a per-flow basis after collecting 100 packets for a given flow to classify. Specifically, flows are identified by means of the standard 5-tuple: src/dst IP, src/dst port, protocol.

The initial choice to use VNFs simplifies the integration of ML workflows into the network infrastructure and provides flexibility by decoupling our approach from specific technologies. This ensures that our method is adaptable to different platforms and can evolve as new technologies emerge.

Within the \underline{C\&OP plane}, an algorithm that computes the optimal deployment of the WL-VNFs is implemented. 

A non-linear integer programming model is used to mathematically model the problem; a Biased Random-Key Genetic Algorithm (BRKGA) \cite{ref:vrp_brkga_3} is instead used to solve it on large-size instances. 
More specifically, the deployment strategy assigns a unique color to each WL-VNF. The problem is defined as a \textit{shortest path problem with coloring constraints}, where a solution is feasible if the computed shortest path ensures that each color (i.e. each deployed WL-VNF) is traversed by the network traffic, as this reconstructs the decomposed SL. Additionally, a solution is considered to be also optimal if it guarantees the minimum shortest path cost. 

Without loss of generality, a sample topology considered for our experiments is shown in Fig. \ref{fig:fig1}. The shown distribution plan of the WL-VNFs has been generated by the optimization algorithm. Each square node in the topology of the figure represents a PDP device which hosts a WL-VNF. Each  WL-VNF instantiated into a PDP device is indicated in the figure with a different color: red, green, and blue; while hosts are colored red or blue depending on whether they are malicious or normal hosts. 

Instead, Fig. \ref{fig:sl-top}, depicts our benchmark solution in which entire SLs are deployed (SL-VNFs represented in red). 

In the case of the WL-VNFs deployment, each packet/flow shall follow a \textit{shortest path that contains at least three different colors}, which means traversing all the WLs that make up the SL. The benchmark  SL-VNFs deployment, instead,  guarantees that each packet/flow follows the \textit{shortest path that includes at least one SL-VNF}.

As regards the emulation of the  \underline{AIePDP plane}, the OpenFlow programmable Open vSwitch (OvS) devices that make up the AIePDP are emulated as Ubuntu 20.04 Virtual Machines (VM) with 2 cores and 4GB of RAM. They are connected to build the network topology in Fig.\ref{fig:fig1} using the Graphical Network Simulator version 3 (GNS3).
Python3 programming language was used to implement the WL/SL-VNFs. The chosen SDN controller is Ryu, also written in Python.

\subsection{Performance Evaluation Results}
The performance metrics considered are: the \textit{average network throughput}  guaranteed when devices are equipped with SL-VNFs/WL-VNFs; the \textit{average RAM and CPU additional utilization} of PDP devices due to the execution of the VNFs; the \textit{detection capability} of the in-network IPS, in terms of time required to compute the inference on the traversing flow (considering both feature extraction and model interrogation). 

The host-to-host benign traffic is generated as follows:
\begin{itemize}
    \item H1-H2: S7-S6-S2-S6
     \item H8-H10: S8-S3-S1-S0
      \item H5-H6: S9-S3-S1-S0-S6
       \item H8-H3: S8-S7-S6-S2
\end{itemize}

Without loss of generality, three levels of traffic loads are considered, characterized by packets of variable size sent for 300 seconds at the rates of 10 pkts/s (TCP), 100 pkts/s (TCP), and 1000 pkts/s (TCP), respectively representing situations of Low, Medium, and High traffic loads. 

The attackers, instead, perform volumetric DoS/DDoS by flooding the network with malicious traffic at a variable pkts/s rate. 
From the analysis conducted on the CIC-DDoS2019 \cite{cicddos} dataset, it results that packet sizes associated with malicious DoS/DDoS traffic are, on average, an order of magnitude larger than those observed in benign traffic. To account for this difference, we vary packet sizes for benign traffic in the range of 68 to 1500 bytes, with 68 bytes being the minimum MTU. In contrast, packet sizes for malicious traffic are uniformly distributed in the range of 68 to 15000 bytes.
The considered volumetric DDoS attacks, treated in the CIC-IDS-2018, namely the Http Unbearable Load King (HULK) DDoS, is considered, which can flood a target with a huge amount of HTTP requests, exhausting its resources.

In initial tests under the assumption that no attacks take place and WL-VNFs executed in the switches, we observe that, if the traffic load increases from low to medium and high, the network throughput decreases by 3.7\%, 7.14\%, and 15\% respectively. As expected, worse results are obtained with an SL-VNF deployment, in which throughput reductions of 25\% about 38\%, and 49\% are observed under small, medium, and high traffic load conditions respectively. 
The benefits of WL-VNFs are not only thanks to the lower-complexity model installed in each device but also to the distributed computation of the features needed to feed the model and get the detection. Indeed, with WL-VNFs, each in-network predictor must extract only 24 features, instead of the 72 features needed when executing the SL-VNF. 

The most relevant and interesting benefits are, however, observable when a DDoS takes place. The considered malicious entities  performing the DDoS are the malicious hosts H9, H11, and H7, whose traffic follows the routes below:
\begin{itemize}
    \item H9-H2: S7-S6-S2-S6
     \item H11-H6: S9-S3-S1-S0-S6
      \item H7-H4: S8-S7-S6-S2  
      \end{itemize}

Taking as reference this experimental setup, and considering the Medium load scenario, we increased the rate of the DDoS attack, starting from 100 TCP pkt/s sent by each of the malicious hosts and reaching 1000 pkt/s, with an increasing step of 100.  
For each attack rate, we conducted a simulation of 300s.

\begin{figure}[!h]
    \centering
    \includegraphics[width=3in]{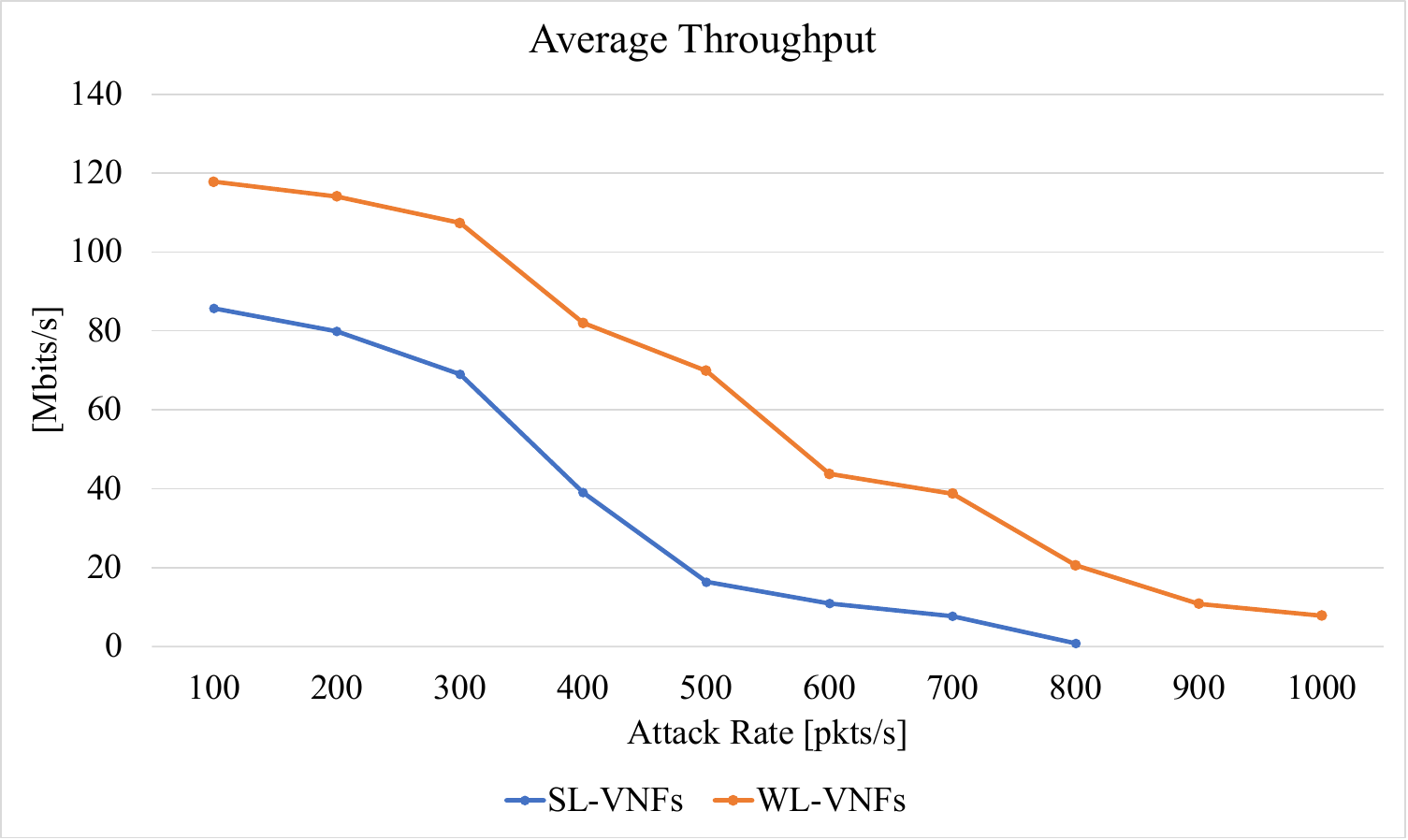}
    \caption{Average Network Throughput: SL-VNFs vs. WL-VNFs Deployment.}
    \label{fig:throughput_attack_rate}
\end{figure}

In Fig. \ref{fig:throughput_attack_rate} results in terms of average network throughput, while increasing the attack rate, are given. 

The throughput decreases with an increase in the amount of traffic generated by the distributed malicious hosts, showing two critical points around 400 pkts/s (for each attacker) and 700 pkts/s, for which there is a drastic throughput reduction. However, it is observed that the WL-VNFs deployment always shows the best gain for the network, in terms of throughput.

This confirms that the higher SL model complexity imposes a too heavy computational overhead on the network devices. As widely expected, instead, with the WL-VNFs deployment the distribution of the computational load is distributed across the network and the PDP devices forwarding capabilities are preserved.

For an attack rate of 400pkt/s, the guaranteed average network throughput with WL-VNFs increases by 63\% with respect to the SL-VNFs setup. Real benefits can be observed when the attack rate increases to 700 pkts/s per attacker. The average throughput of the network drastically lowers ($\sim$ 6 Mbits/s). In such a case, the WL-VNFs deployment setting makes the switches able to handle network traffic with an average throughput $\sim$ 40Mbits/s. Increasing the attack rate to 800pkts/s, the data plane devices running the SL-VNFs are not able to handle the network traffic while classifying it, and the average network throughput is almost zeroed out. The WL-VNFs deployment setting, instead, does not lead the data plane devices to be torn down, continuing to provide network forwarding capabilities as well as detection capabilities with a minimum average throughput rate of about 10Mbits/s under extreme attack conditions.

\begin{figure}[!h]
    \centering
    \includegraphics[width=3in]{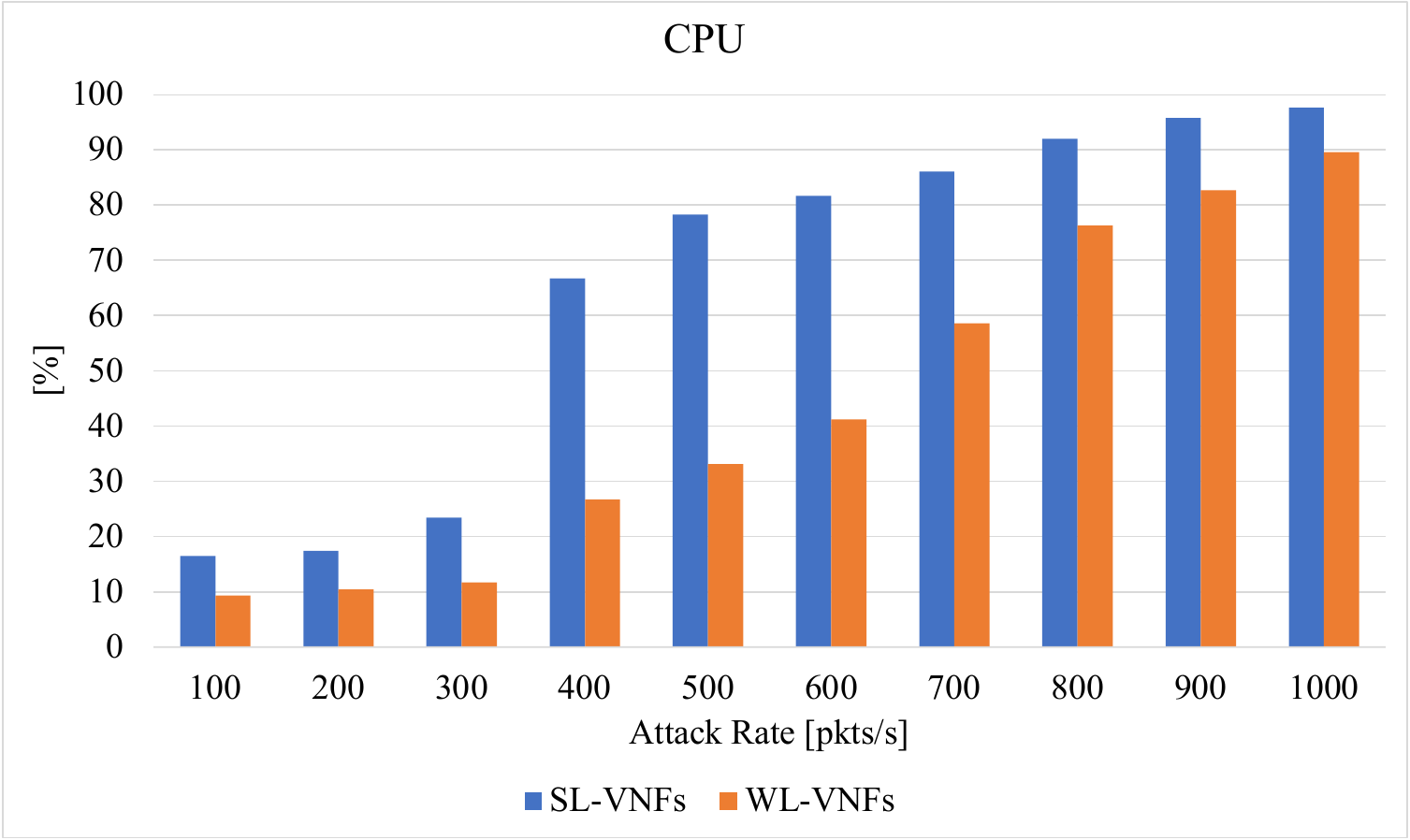}
    \caption{Average PDPs' CPU Utilization: SL-VNFs vs. WL-VNFs Deployment.}
    \label{fig:cpu}
\end{figure}

Tests are also conducted on the network forwarding elements’ resource overhead, by measuring the average CPU and RAM consumption on the devices running the models (WL-VNFs/SL-VNFs). 
Only the results referring to the CPU are shown in Fig. \ref{fig:cpu} for space reasons, but those relating to Memory are similar. 

When using SL-VNFs, the additional overhead causes a resource consumption of approximately $\sim$80\% when the attack rate is 400-500pkt/s. When the attack rate increases to 700-800pkts/s, the CPU/RAM consumption reaches $\sim$91\% and it is almost 100\% for a 900-1000 pkts/s attack rate. Indeed, at this point, the average network throughput (as previously demonstrated) is zeroed out.

With WL-VNFs, instead, the switches’ resource consumption is preserved, maintaining an average CPU/RAM utilization of about 50-60\% when the attack rate is 600-700 pkts/s, with a gain of about 50\% with respect to the SL-VNFs deployment setting.
This demonstrates that leveraging WL-VNFs enables the network to scale better with the growing amount of traffic and withstand massive attacks. 

\begin{figure}[!h]
    \centering
    \includegraphics[width=3in]{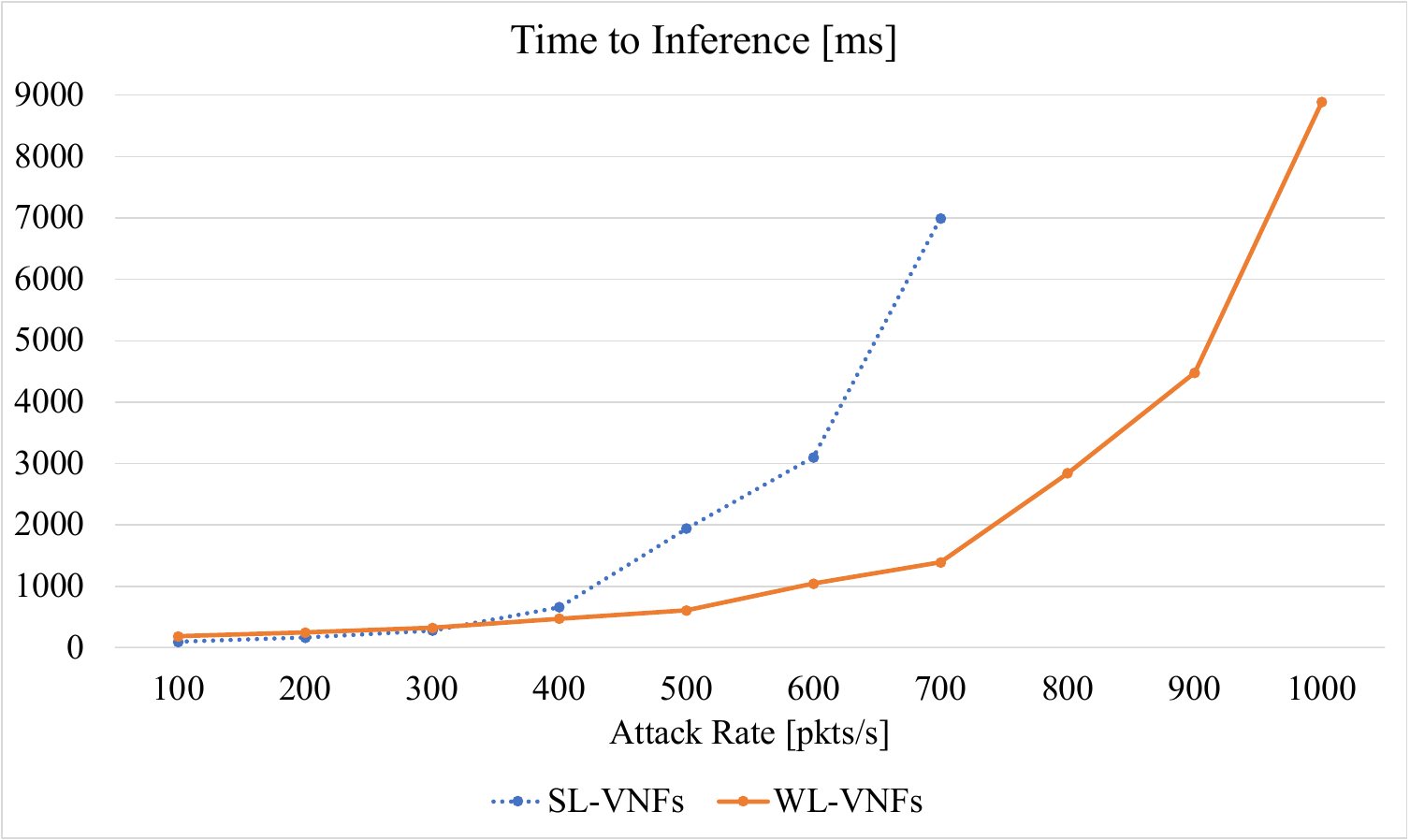}
    \caption{Time To Inference: SL-VNFs vs. WL-VNFs Deployment.}
    \label{fig:time_to_inference}
\end{figure}

The last curves in Fig. \ref{fig:time_to_inference} show the time needed to classify flows with WL-VNFs and with SL-VNFs when traffic increases. It can be observed that as long as each attacker generates up to 200-300 pkt/s of malicious traffic, the SL-VNFs can compute the final inference in less time than the WL-VNFs (about 16\% when the attack packet rate is 300 pkt/s). 

This is due to the typical feature of the WL-VNF implementation of having to pass intermediate inferences to other predictors in the network to perform the final detection. However, when the attack rate increases and the switches are overwhelmed with network packets to analyze, this advantage disappears and the WL-VNF setup clearly shows its benefits.

More precisely, when the DDoS takes place with 700pkts/s per attacker, the WL-VNFs can perform detection in about $\sim$1400ms. 
In the same setting, the SL-VNFs show a time to compute a single detection of about $\sim$7000 ms. Things get even worse when the attack rate increases: after 800pkts/s per attacker the SL-VNFs are unable to respond in a reasonable amount of time until it is completely torn down due to the huge amount of packets to handle. WL-VNFs, instead, can maintain a reasonable time-to-detection with a maximum peak of about $\sim$ 9000 ms.  

Finally, the additional overhead due to the fact that packets do not follow the shortest path to the destination is calculated using an ad hoc metric, which considers the difference between the colored path and the shortest path for each pair of source-destination nodes.
Multiple tests have certified that the overhead varies with the size of the network while it remains almost constant when the density varies. In the tests performed, however, it was observed that the calculated overhead percentage always assumes widely acceptable values (around an average value of 3.5 \% in the worst case scenario).

These results shed light on the importance of considering model-splitting techniques such as in the proposed paradigm of ``projecting ensemble learning on the network".

\section{Distributed In-network Defense: Future Research Directions}
After assessing the potential of a programmable and ML-enabled data plane of a future 6G network to implement an in-network distributed IDS/IPS paradigm, we now describe some research challenges that the new paradigm opens and that need to be effectively addressed. \\

\textit{Dynamic Deployment Strategy}.
One of the key challenges, common to any distributed inference scenario, is to constantly optimize the  deployment of the addressed WLs in-network. Therefore, in the future, a new class of efficient WL-VNF deployment/chaining strategies shall be studied, capable of dynamically deciding the most effective WL-VNF re-distribution whenever changes in traffic conditions or topology occur (due to a link failure, for example), while being able to classify all traffic traversing their minimum-cost paths from source to destination and minimizing the load on network elements. \\

\textit{Hardware and Programming Languages Limitations}.
ML-DL tasks rely on mathematical and algebraic computations involving different data types, especially floating-point numbers, which might degrade forwarding activities. So, at the switch level these are not allowed and programming languages for PDP are characterized by a limited set of operations, this posing a daunting technological barrier. The seamless integration of new paradigms, like the one proposed, into next-generation networks will therefore have to pass through an enhancement of computational fabrics to support a wider set of operations and data types. \\

\textit{Lightweight ML functions}.
ML/DL learning models are known to be resource-hungry, both in terms of memory and computational capacity. Therefore, they need to be rethought in order to find a proper trade-off between model complexity and accuracy and thus to make them deployable on constrained (especially in terms of storage) forwarding devices. Some advancements in this aspect have been made by resorting to well-established model quantization techniques, like Binary Neural Networks (BNN). However, new ML/DL models specifically designed and suited for networking devices should be developed to make the network capable of accelerating highly distributed AI-related tasks. \\

\textit{Near Zero-Latency traffic Detection.} 
New PDP devices promise an unprecedented magnitude of line-rate packet forwarding (ranging from FPGA sNIC with a line-rate of 100Gb/s to Barefoot Tofino that allows for 12.8Tb/s line-rate processing). Despite this, to make in-network intrusion detection deeply pervasive in future networks, the PDP devices should no longer be conceived to keep high throughput in packet forwarding as the only primary goal. Rather, they must also be designed and deployed to support computation, like the one required for the purpose of the presented research. 

\section{Conclusion}
In this paper, a disruptive paradigm has been introduced aimed at defining a new approach to intrusion detection in which very lightweight learning models, resulting from splitting a strong model, are dynamically distributed and appropriately chained within the switches of a next-generation PDP.
The objective is to have future 6G networks natively protected from malicious traffic by the same devices in their data plane, and at the same time to avoid overloading the forwarding functions with heavy learning models implemented in the switches. 
An initial set of measurements was shown to provide a proof-of-concept of the new paradigm and to highlight its potential.
Finally, indications were provided on the research path ahead for the full realization of the distributed in-network IPS hypothesized in this paper.

\bibliographystyle{IEEEtran}
\bibliography{biblio}

\begin{thebibliography}{10}
\providecommand{\url}[1]{#1}
\csname url@samestyle\endcsname
\providecommand{\newblock}{\relax}
\providecommand{\bibinfo}[2]{#2}
\providecommand{\BIBentrySTDinterwordspacing}{\spaceskip=0pt\relax}
\providecommand{\BIBentryALTinterwordstretchfactor}{4}
\providecommand{\BIBentryALTinterwordspacing}{\spaceskip=\fontdimen2\font plus
\BIBentryALTinterwordstretchfactor\fontdimen3\font minus \fontdimen4\font\relax}
\providecommand{\BIBforeignlanguage}[2]{{%
\expandafter\ifx\csname l@#1\endcsname\relax
\typeout{** WARNING: IEEEtran.bst: No hyphenation pattern has been}%
\typeout{** loaded for the language `#1'. Using the pattern for}%
\typeout{** the default language instead.}%
\else
\language=\csname l@#1\endcsname
\fi
#2}}
\providecommand{\BIBdecl}{\relax}
\BIBdecl

\bibitem{Park23}
C.~Park, K.~Park, J.~Song, and J.~Kim, ``Distributed learning-based intrusion detection in 5g and beyond networks,'' in \emph{Proceedings of the 2023 European Conference on Networks and Communications \& 6G Summit (EuCNC/6G Summit)}, 2023, pp. 490--495.

\bibitem{bosshart2014p4}
P.~Bosshart \emph{et~al.}, ``P4: Programming protocol-independent packet processors,'' \emph{ACM SIGCOMM Computer Communication Review}, vol.~44, no.~3, pp. 87--95, 2014.

\bibitem{Cordero}
W.~L. da~Costa~Cordeiro, J.~A. Marques, and L.~P. Gaspary, ``Data plane programmability beyond openflow: Opportunities and challenges for network and service operations and management,'' \emph{Jour. of Netw Syst Management (2017) 25:784–818}, vol.~25, pp. 784--818, 2017.

\bibitem{Tarik23}
S.~Kianpisheh and T.~Taleb, ``A survey on in-network computing: Programmable data plane and technology specific applications,'' \emph{IEEE COMMUNICATIONS SURVEYS \& TUTORIALS}, vol.~25, no.~1, pp. 701--761, 2023.

\bibitem{Capone}
D.~Moro, G.~Verticale, and A.~Capone, ``Network function decomposition and offloading on heterogeneous networks with programmable data planes,'' \emph{IEEE Open Journal of the Communications Society}, vol.~2, pp. 1874--1885, 2021.

\bibitem{MAHM23}
S.~Schwarzmann \emph{et~al.}, ``An intelligent user plane to support in-network computing in 6g networks,'' in \emph{Proceedings of the IEEE International Conference on Communications (ICC)}, 2023.

\bibitem{Gao}
Y.~Gao and Z.~Wang, ``{A Review of P4 Programmable Data Planes for Network Security},'' \emph{Mobile Information Systems.}, Nov. 2021.

\bibitem{saquetti2021toward}
M.~Saquetti \emph{et~al.}, ``Toward in-network intelligence: Running distributed artificial neural networks in the data plane,'' \emph{IEEE Communications Letters}, vol.~25, no.~11, pp. 3551--3555, 2021.

\bibitem{xavier2021programmable}
B.~M. Xavier, R.~S. Guimarães, G.~Comarela, and M.~Martinello, ``Programmable switches for in-networking classification,'' in \emph{IEEE INFOCOM 2021-IEEE Conference on Computer Communications}.\hskip 1em plus 0.5em minus 0.4em\relax IEEE, 2021, pp. 1--10.

\bibitem{Busse}
C.~Busse-Grawitz \emph{et~al.}, ``{pforest: In-network inference with random forests},'' \emph{arXiv preprint arXiv:1909.05680}, 2019.

\bibitem{Zheng}
C.~Zheng \emph{et~al.}, ``{IIsy: Practical In-Network Classification},'' \emph{arXiv preprint arXiv:2205.08243}, 2022.

\bibitem{Lee2020Nov}
J.-H. Lee and K.~Singh, ``{SwitchTree: in-network computing and traffic analyses with Random Forests},'' \emph{Neural Comput. {\&}. Applic.}, pp. 1--12, Nov. 2020.

\bibitem{NetBeacon}
\BIBentryALTinterwordspacing
G.~Zhou, Z.~Liu, C.~Fu, Q.~Li, and K.~Xu, ``An efficient design of intelligent network data plane,'' in \emph{32nd USENIX Security Symposium (USENIX Security 23)}.\hskip 1em plus 0.5em minus 0.4em\relax Anaheim, CA: USENIX Association, Aug. 2023, pp. 6203--6220. [Online]. Available: \url{https://www.usenix.org/conference/usenixsecurity23/presentation/zhou-guangmeng}
\BIBentrySTDinterwordspacing

\bibitem{ref:vrp_brkga_3}
P.~Festa, F.~Guerriero, M.~G. Resende, and E.~Scalzo, ``A {BRKGA} with implicit path-relinking for the vehicle routing problem with occasional drivers and time windows,'' in \emph{Lecture Notes in Computer Science – MIC 2022: 14th Metaheuristics International Conference, forthcoming.}, 2023.

\bibitem{cicddos}
I.~Sharafaldin, A.~H. Lashkari, S.~Hakak, and A.~A. Ghorbani, ``{Developing Realistic Distributed Denial of Service (DDoS) Attack Dataset and Taxonomy},'' in \emph{{2019 International Carnahan Conference on Security Technology (ICCST)}}.\hskip 1em plus 0.5em minus 0.4em\relax IEEE, 2019, pp. 01--03.

\end{thebibliography}

\begin{IEEEbiographynophoto}{Mattia Giovanni Spina} is a PhD student at the University of Calabria (Italy). His research interest is in the area of security in future generation networks and distributed AI in-network architectures.
\end{IEEEbiographynophoto}
\vspace{-1.5 cm} 

\begin{IEEEbiographynophoto}{Floriano De Rango} 
is associate professor of Telecommunications at the University of Calabria (Italy). His research interests include security in wireless and IoT networks and networking solutions for V2X systems.
\end{IEEEbiographynophoto}
\vspace{-1.5 cm} 

\begin{IEEEbiographynophoto}{Antonio Iera} 
is full professor of Telecommunications at the University of Calabria (Italy). His research interests include next generation mobile and wireless networks and the Internet of Things.
\end{IEEEbiographynophoto}
\vspace{-1.5 cm}

\begin{IEEEbiographynophoto}{Edoardo Scalzo}
is a junior researcher of Operations Research at the University of Calabria (Italy). His research interests include network optimization, logistics and combinatorial optimization.
\end{IEEEbiographynophoto}
\vspace{-1.5 cm}

\begin{IEEEbiographynophoto}{Francesca Guerriero}
is a full professor of Operations Research at the University of Calabria, Italy. Her primary research interests revolve around network optimization, logistics, combinatorial optimization, and the intersection of optimization and big data.
\end{IEEEbiographynophoto}
\vspace{-1.5 cm}

\end{document}